\documentclass[aps,pra,twocolumn,superscriptaddress,floatfix]{revtex4-2}
\usepackage{multirow}
\usepackage{graphicx}
\usepackage{siunitx}
\usepackage{hyperref}
\usepackage[english]{babel}
\usepackage{braket}
\usepackage{nicefrac}
\newcommand{\potassium}{$^{39}$K}

\usepackage{amsmath}
\usepackage{graphicx}
\usepackage{siunitx}
\usepackage{booktabs}
\usepackage{enumerate}

\graphicspath{{images/}}

\usepackage[rgb]{xcolor}

\newcommand{\abs}[1]{\left|#1\right|}

\newcommand{\molket}[4]{$\ket{#1,#2,#3}\ket{#4}$}
\newcommand{\groundket}[1]{$\ket{#1}\ket{#1}$}

\addto\extrasenglish{}

\def\equationautorefname#1#2\null{Eq.#1(#2\null)}

\begin{document}


\title{Photodissociation of long-range Rydberg molecules}


\author{Michael Peper}
\affiliation{Laboratory of Physical Chemistry, ETH Z\"urich, 8093 Z\"urich, Switzerland}
\author{Johannes Deiglmayr}
\affiliation{Laboratory of Physical Chemistry, ETH Z\"urich, 8093 Z\"urich, Switzerland}
\affiliation{Department of Physics and Geoscience, University of Leipzig, 04109 Leipzig, Germany}
\email{johannes.deiglmayr@uni-leipzig.de}


\date{\today}

\begin{abstract}
We present photodissociation of ultracold long-range Rydberg molecules as a tool to characterize their electronic properties. We photoassociate $^{39}$K$_2$ $37\,^2{\rm P}$ molecules with highly entangled electronic and nuclear spins of the two bound atoms and quantify the entanglement by projection of the molecular state onto non-interacting atoms using radiofrequency photodissociation. By comparison of experimental photodissociation rates with theoretical predictions we further characterize the electronic and nuclear wavefunction of the photoassociated molecules. Based on the complete characterization of the formed long-range Rydberg molecules, we demonstrate a full hyperfine-spin flip of a free ground-state atom through the interaction with a Rydberg atom.
\end{abstract}


\maketitle

\section{Introduction}
Photodissociation of cold molecules is a powerful tool, providing quantum control of the photofragments' internal and translational state \cite{chandlerQuantumControlLightinduced2016,mcdonaldPhotodissociationUltracoldDiatomic2016,zhouSecondScaleCoherenceMeasured2020}. It has been used, \textit{e.g.}, to produce atoms and molecules with very low kinetic energy for high-resolution collision and spectroscopic studies \cite{matthewsFullyStateselectedVMI2007,sukzhaoSlowMoleculesProduced2009} and aligned molecular samples  \cite{dehmelt62a,richardson68a}, as well as beams of spin-polarized atoms~\cite{rakitzisSpinPolarizedHydrogenAtoms2003,sofikitisHighlyNuclearSpinPolarizedDeuterium2017}. Photodissociation of very weakly bound ultracold molecules was used to probe interactions in highly-degenerate quantum gases \cite{regalCreationUltracoldMolecules2003,greinerDetectionSpatialCorrelations2004,chinObservationPairingGap2004,schunckDeterminationFermionPair2008}. Here we apply photodissociation to long-range Rydberg molecules (LRMs). LRMs are bound states of a ground-state atom located within the orbit of a Rydberg electron, where the binding is provided by the Rydberg-electron--ground-state atom scattering interaction \cite{greeneCreationPolarNonpolar2000}. LRMs have previously been studied by photoassociation spectroscopy \cite{lettSpectroscopyNaPhotoassociation1993,bendkowsky09a,shaffer2018ultracold} and radiofrequency (RF) spectroscopy \cite{peper2020heteronuclear}, extracting, \textit{e.g.}, electron-atom scattering parameters from binding energies \cite{sassmannshausenprl2015,maclennanDeeplyBound24D2019,engelPrecisionSpectroscopyNegativeIon2019} and determining dipole moments \cite{li2011,booth2015}.

Here we present photodissociation as a probe of electronic properties of LRMs. We find that the binding of a LRM can be effectively switched off by transferring the Rydberg electron into an orbit where it does not interact with the ground-state atom, thus projecting the molecular electronic state onto an unperturbed basis of atomic product states where the quantum numbers of Rydberg atom and ground-state atom are well defined (\autoref{fig:levelscheme}). As a first application, we demonstrate threshold photodissociation of LRMs of \potassium$_2$ with highly entangled spins of the Rydberg and ground-state atom, formed by photoassociation~\cite{niederprumRydbergMoleculeInducedRemote2016a}. This allows for ($i$) the manipulation of the internal state of the ground-state atom through interaction with the Rydberg atom at large interatomic separations, and ($ii$) a tomographic characterization of the electronic structure of the photoassociated LRMs.

\section{Electronic properties of long-range Rydberg molecules}

\begin{figure}[bt!]
	\centering
    \includegraphics[width=\linewidth]{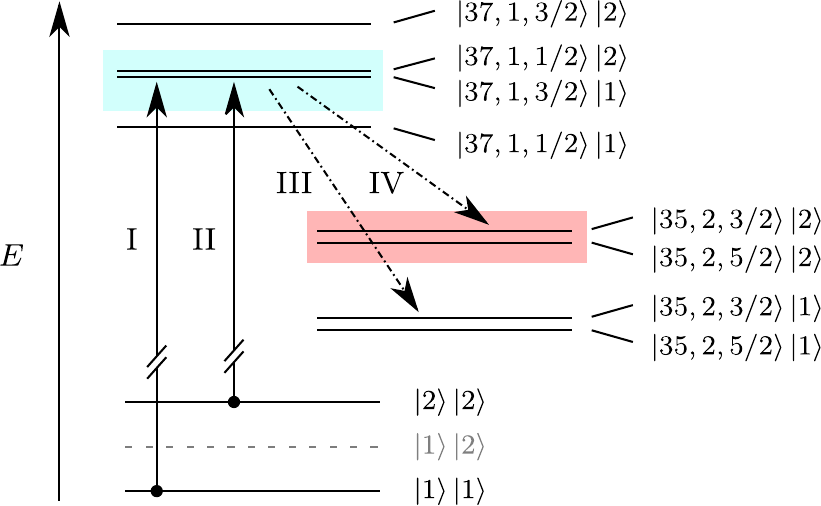}
	\caption{\label{fig:levelscheme}Energy-level scheme (not to scale) of the relevant molecular asymptotes involved in the photoassociation and photodissociation of \molket{37}{1}{j}{F} LRMs. Arrows I and II indicate photoassociation from ground-state atoms prepared in $F=1$ and $F=2$, respectively. The dashed-dotted arrows III and IV indicate RF photodissociation of the LRMs to \molket{35}{2}{j'}{F'}. The potential-energy curves in the regions shaded in light blue and light red are depicted in Figs.~\ref{fig:P_potentialandspectra}\,a) and \ref{fig:D_potentialandspectra}\,a), respectively. State labels \molket{n}{l}{j}{F} are defined in the text.}
\end{figure}

The electronic Hamiltonian of a LRM in the Born-Oppenheimer approximation (neglecting the hyperfine interaction of the Rydberg atom) is given by
\begin{equation}\label{eq:Chp5:SimpleHamiltonian}
    H=H_{0}+ A_{\rm{hfs}}\,\vec{s}_{\rm gs}\cdot\vec{i}_{\rm gs} +A_{\rm{so}}\,\vec{l}\cdot\vec{s}_{\rm Ryd} + \Delta V\,\vec{s}_{\rm Ryd}\cdot\vec{s}_{\rm gs} + \bar{V},
\end{equation}
which includes $H_0$, the Coulomb interaction of the ion core with the Rydberg electron \cite{peperPrecisionMeasurementIonization2019}, the spin-orbit coupling of the Rydberg-electron spin $\vec{s}_\mathrm{Ryd}$ and orbital angular momentum $\vec{l}$ to form $\vec{j}$ with a state-dependent constant $A_\mathrm{so}$, and the hyperfine coupling between the electronic ($\vec{s}_\mathrm{gs}$) and nuclear ($\vec{i}_\mathrm{gs}$) spin of the ground-state atom to form $\vec{F}$ with the coupling constant $A_\mathrm{hfs}$. The binding Fermi contact interaction $V_\mathrm{c}$ between the Rydberg and the ground-state atom depends on the total electronic spin $S$ of the collision complex \cite{Anderson2014a,sassmannshausenprl2015,marksonTheoryUltralongRangeRydberg2016,eilesHamiltonianInclusionSpin2017a}. In Eq.~\eqref{eq:Chp5:SimpleHamiltonian}, $V_\mathrm{c}$ is split into a spin-dependent part $\Delta V=V_\mathrm{c}^\mathrm{3}-V_\mathrm{c}^\mathrm{1}$ and a spin-independent part $\bar{V}=\frac{3}{4}V_\mathrm{c}^\mathrm{3}+\frac{1}{4}V_\mathrm{c}^\mathrm{1}$, where superscripts 1 and 3 refer to the scattering in singlet and triplet channels, respectively. The spin-dependent term $\Delta V\,\vec{s}_{\rm Ryd}\cdot\vec{s}_{\rm gs}$ takes the form of the Heisenberg-Dirac Hamiltonian for the two-electron exchange interaction~\cite{diracQuantumMechanicsManyelectron1929}, which creates entanglement of the hyperfine state of the ground-state atom $F$ and the spin-orbit state $j$ of the Rydberg atom~\cite{niederprumRydbergMoleculeInducedRemote2016a}. At $n=37$, the fine-structure splitting of the \potassium{} $n\,^2{\rm P}_j$ Rydberg state is almost degenerate with the ground-state hyperfine splitting of $^{39}$K (\SI{456.2}{MHz} and \SI{461.7}{MHz}, respectively)~\cite{peperPrecisionMeasurementIonization2019}, and the entanglement may become especially large (see \autoref{fig:levelscheme}).

The Hamiltonian~\eqref{eq:Chp5:SimpleHamiltonian} conserves the quantity
\begin{equation}
\Omega=m_l+m_{s_\mathrm{Ryd}}+m_{s_\mathrm{gs}}+m_i = m_j + m_F,
\end{equation}
which is the projection of the total angular momentum onto the internuclear axis (chosen along the $z$-axis), and is commonly represented in a basis comprising direct-product states of the Rydberg atom and ground-state atom {\molket{n}{l}{j,m_j}{F,m_F}} ~\cite{marksonTheoryUltralongRangeRydberg2016,eilesHamiltonianInclusionSpin2017a}. Here, $\ket{n,l,j,m_j}$ denotes the state of the Rydberg atom and $\ket{F,m_F}$ denotes the hyperfine state of the $4\,^2\mathrm{S}_{1/2}$ ground state of \potassium{}. For clarity we will omit $m_j$ and $m_F$ in the following.  As commonly done in the treatment of LRMs, we consider Rydberg and ground-state atom as distinguishable particles~\cite{greeneCreationPolarNonpolar2000,li2011}.

\begin{figure}[tb!]
	\centering
	\includegraphics[width=\linewidth]{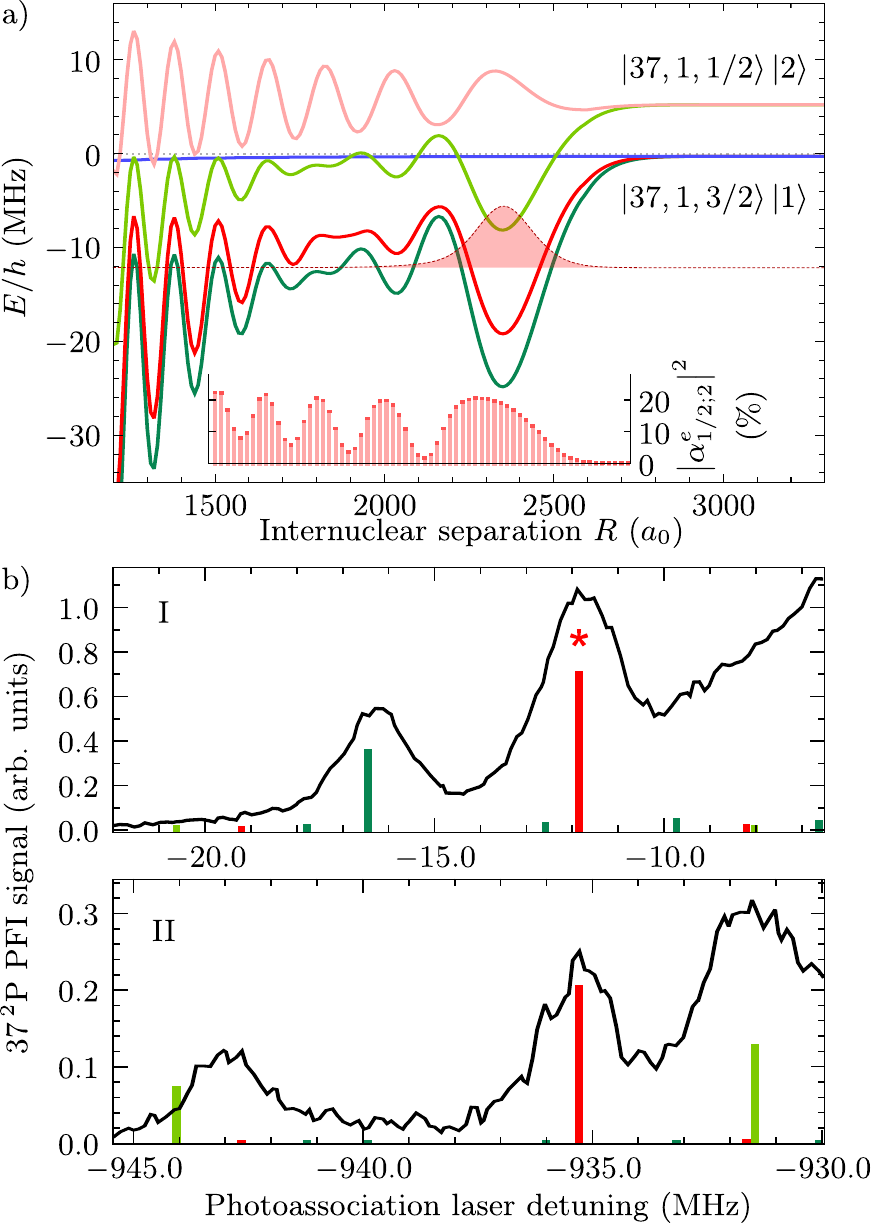}
	\caption{\label{fig:P_potentialandspectra}a) PECs of states associated with the near-degenerate \molket{37}{1}{3/2}{1} and \molket{37}{1}{1/2}{2} asymptotes. Shown are all states for $\Omega=1/2$ with $^3\Sigma^+$ (dark and light green), $^{1,3}\Sigma^+$ (red), and $^{1,3}\Pi$ (blue) symmetry. The red-filled curve depicts the vibrational wavefunction of the vibrational ground state in the outermost well of the $^{1,3}\Sigma^+$ state correlated to the \molket{37}{1}{3/2}{1} asymptote, $\Psi_e$. Its $R$-dependent absolute square of the coefficient $\alpha_{1/2;2}^e$ is depicted in the inset. b) photoassociation spectra recorded close to the \molket{37}{1}{3/2}{1} asymptote after preparing the ground-state atoms in (I) the $F=1$ and (II) the $F=2$ hyperfine level. For measurement II, the intensity of the photoassociation laser was increased by a factor of 4.7(1) to improve the signal-to-noise ratio. The laser frequency is given relative to the frequency of the atomic \molket{37}{1}{3/2}{1}$\leftarrow$\groundket{1} transition. Solid bars indicate the calculated positions and strengths of photoassociation resonances (color coding as in a)), as described in Appendix \ref{app:calcPA}. The red star marks the resonance probed by photodissociation (\autoref{fig:D_potentialandspectra}).}
\end{figure}

\autoref{fig:P_potentialandspectra}\,a) depicts calculated potential-energy curves (PECs) for $\Omega=1/2$ states correlated to the nearly degenerate \molket{37}{1}{j}{F} asymptotes. The calculations follow Refs.~\cite{eilesHamiltonianInclusionSpin2017a,eilesFormationLongrangeRydberg2018,peperFormationUltracoldIon2020a,peper2020heteronuclear} with \textit{ab-initio} calculated scattering phase shifts~\cite{eilesFormationLongrangeRydberg2018} and a computational basis including the Rydberg state of interest, all Rydberg states of the four closest hydrogenic manifolds (two above and two below the target state) and the low-$l$ Rydberg states within this range, as well as all hyperfine states $\ket{F,m_F}$ of the ground state atom. Because the PECs are degenerate in $\Omega$~\cite{eilesHamiltonianInclusionSpin2017a}, we label electronic states by approximate term symbols $^3\Sigma^+$ (dark and light green), $^{1,3}\Sigma^+$ (dark and light red), and $\Pi$ (blue), where the superscript indicates if the binding results only from triplet (3) or singlet-triplet mixed scattering (1,3). Only in $^3\Sigma^+$ states, $F$ (the hyperfine state of the ground-state atom) is a good quantum number, whereas it is mixed in $^{1,3}\Sigma^+$ states~\cite{Anderson2014a}. $\Sigma$ and $\Pi$ refer to states with and without $m_l=0$ contributions, respectively.

\emph{Entangled states}
The states $\ket{\phi}$ depicted in \autoref{fig:P_potentialandspectra}\,a)  exhibit almost exclusively \molket{37}{1}{3/2}{1} and \molket{37}{1}{1/2}{2} character. Their state vectors can thus be approximated by
\begin{equation}
\label{eq:HFS:twostatemodel}
\ket{\phi}=\sqrt{\frac{1}{2}}\left(\alpha_{3/2;1}^\phi\ket{37,1,3/2}\ket{1}+\alpha_{1/2;2}^\phi\ket{37,1,1/2}\ket{2}\right),
\end{equation}
with $R$-dependent mixing coefficients $\alpha_{3/2;1}^\phi$ and $\alpha_{1/2;2}^\phi$.
For the lower $^{1,3}\Sigma^+$ state correlated to the \molket{37}{1}{3/2}{1} asymptote (dark red curve in \autoref{fig:P_potentialandspectra}\,a)), the absolute square of the coefficient $\alpha_{1/2;2}^e$ reaches up to \SI{20}{\percent} (see inset of \autoref{fig:P_potentialandspectra} a)), indicating that the hyperfine state of the ground-state atom and the spin-orbit state of the Rydberg atom are strongly entangled. The oscillatory behaviour of the coefficient is a tell-tale sign that the entanglement is a result of the Fermi contact interaction. We focus on the $v=0$ level of this entangled state, which we refer to as $\Psi_e$ in the following.

\section{Experiments}

\emph{Photoassociation of long-range Rydberg molecules}
Experimentally, the entanglement is observed in photoassociation spectra recorded close to the \molket{37}{1}{3/2}{1} asymptote, depicted in \autoref{fig:P_potentialandspectra}\,b). Samples of ultracold \potassium{} ground-state atoms released from a magneto-optical trap ($T\approx\SI{20}{\micro\kelvin}$, $\rho=\SI{2e10}{\per\cubic\cm}$) are prepared in a single hyperfine state. Subsequently, \potassium$_2$ LRMs are formed by one-photon photoassociation using UV-laser pulses of \SI{30}{\micro\second} length and detected state-selectively on an ion detector by pulsed-field ionization \cite{peper2020heteronuclear}. To probe transitions to the same molecular states when preparing the ground-state atoms in the $F=2$ (II) ground-state hyperfine state, as opposed to $F=1$ (I), the laser frequency has to be reduced by twice the ground-state hyperfine splitting (\autoref{fig:levelscheme}). In both spectra, a strong resonance is observed at a detuning of \SI{-12}{\mega\hertz} (\SI{-935.4}{\mega\hertz}), which is assigned to photoassociation of molecules in state $\Psi_e$. The fact that this resonance is present in both spectra is a clear sign of the mixed $F$ character of this electronic state \cite{niederprumRydbergMoleculeInducedRemote2016a}. From the photoassociation rates normalized to the photoassociation laser intensity and the experimentally determined oscillator strength ratio $f_{j=3/2}/f_{j=1/2}=\num{3.3(5)}$ of the atomic $n\,^2\mathrm{P}_{j}\leftarrow4\,^2\mathrm{S}_{1/2}$ transition, the admixture of \molket{37}{1}{1/2}{2} is found to be $|\alpha_{1/2;2}^e|_\mathrm{exp}^2=\num{0.17(1)}$, which agrees well with the calculated expectation value of $|\alpha_{1/2;2}^e|_\mathrm{theo}^2=\num{0.168}$ for $v=0$. Additional resonances are assigned to levels in the $^{3}\Sigma^+$ states correlated to the \molket{37}{1}{3/2}{1} and \molket{37}{1}{1/2}{2} asymptotes. Because in $^{3}\Sigma^+$ states $F$ is conserved, the former (latter) are only observed when preparing the ground-state atoms in $F=1$ ($F=2$).

\begin{figure}[tb!]
	\centering
	\includegraphics[width=\linewidth]{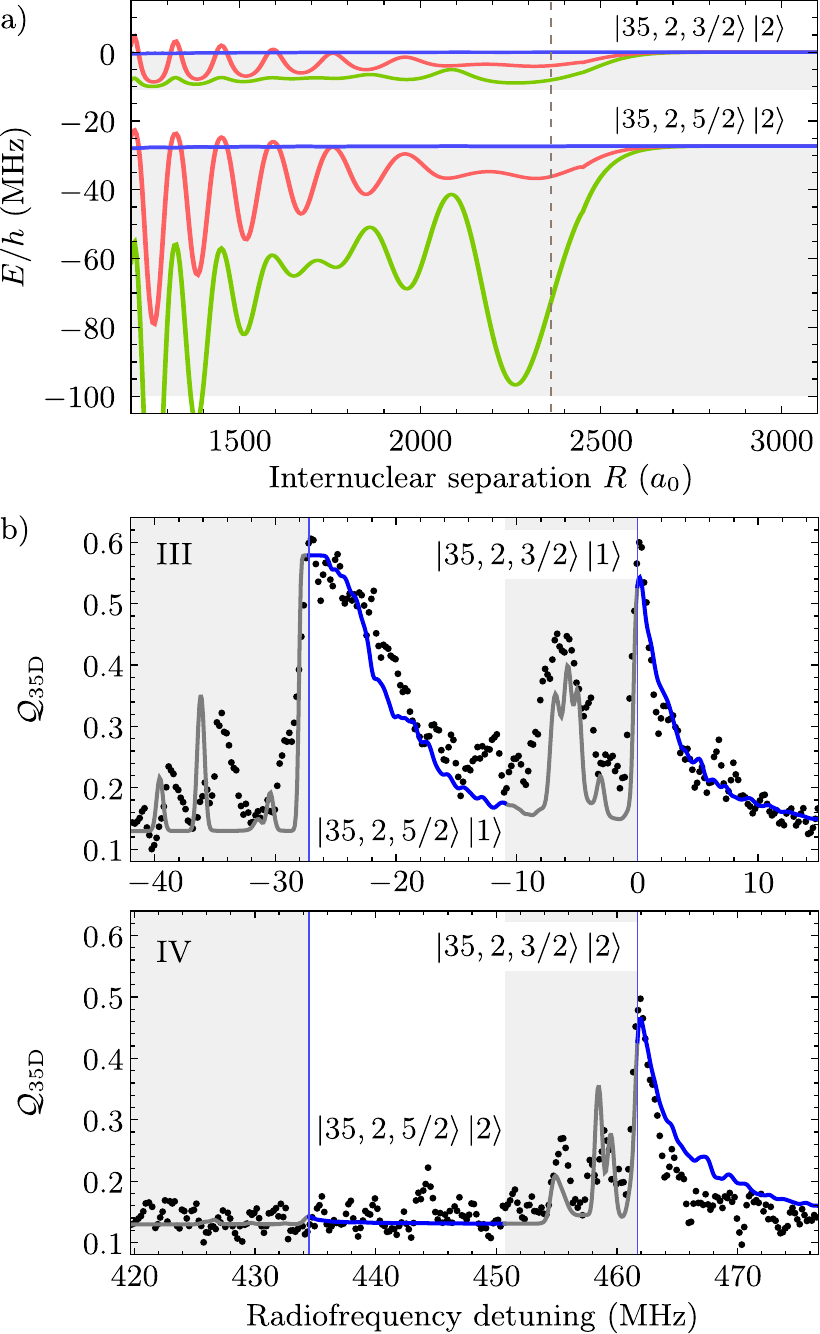}
	\caption{\label{fig:D_potentialandspectra}a) PECs ($\Omega=1/2$) correlated to the \molket{35}{2}{j'}{2} asymptotes. States with $^{3}\Sigma^+$, $^{1,3}\Sigma^+$ and $\Pi$ symmetry are depicted by light green, red and blue lines, respectively. The spectral region in which bound states can exist is highlighted in gray (see also \autoref{fig:D_potentialandspectra}b)). The gray dashed line marks the equilibrium distance of $\Psi_e$. b) RF spectra of $^{1,3}\Sigma^+$ K$_2$ molecules (UV-laser detuning marked by a red star in \autoref{fig:P_potentialandspectra}b)) in the vicinity of the transition to the spin-orbit-split  (III) \molket{35}{2}{j'}{1} and (IV) \molket{35}{2}{j'}{2} asymptotes after a \SI{30}{\micro\second} photoassociation pulse followed by a \SI{10}{\micro\second} RF pulse. Zero RF detuning is set to \SI{86.667}{GHz}, \SI{11.85}{MHz} below the atomic transition $35\,^2\mathrm{D}_{3/2} \leftarrow 37\,^2\mathrm{P}_{3/2}$, corresponding to the transition from the bound molecular state to the dissociation continuum associated with the \molket{35}{2}{3/2}{1} asymptote. The gray-shaded areas mark spectral regions where transitions to bound states are possible (see a)). The vertical thin blue lines mark the transition frequencies from the initial molecular state to the indicated dissociation asymptotes. The gray and blue lines  show the simulated spectrum, where the intensities of the bound (gray) and continuum (blue) parts have been scaled separately to the experimental data because of the different normalization conditions.}
\end{figure}

\emph{Photodissociation} Using RF radiation, we probe the electronic character of the formed LRMs by photodissociation. To this end we photoassociate molecules at the resonance assigned to $\Psi_e$ and drive RF transitions \cite{peper2020heteronuclear,yuMicrowaveTransitionsPairs2013} to molecular states close to the \molket{35}{2}{j'}{F'} asymptotes. The spin-orbit splitting of the $35\,^2\mathrm{D}_j$ state (with an inverted fine-structure splitting of \SI{27.2}{MHz}) is much smaller than the ground-state hyperfine splitting (\SI{461.7}{MHz}) and the four dissociation asymptotes resulting from the combination of $j'=3/2$, $j'=5/2$, $F'=1$, and $F'=2$ group into two fine-structure doublets (\autoref{fig:levelscheme}). For the relevant internuclear separations, the electronic states correlated to \molket{35}{2}{j'}{F'} have an almost pure ground-state hyperfine character, the admixture from the other ground-state hyperfine character being less than \SI{0.1}{\percent}. The potential energy curves of the states correlated to the \molket{35}{2}{j}{2} asymptotes are depicted in \autoref{fig:D_potentialandspectra}\,a). The $\Sigma^+$ states correlated to these asymptotes have mixed $j$ character \cite{Anderson2014}, creating a spectral window above the $j=5/2$ asymptote where no bound states exist. This gap in the bound-state spectrum allows us to study undisturbed transitions into the continuum above each asymptote.

Pulsed RF radiation ($\sim$\SI{86}{GHz}) is applied to the LRMs by sextupling the output of a RF generator \cite{peperPrecisionMeasurementIonization2019}, coupling it to free space via a matched horn and sending it into the chamber through a quartz view port. The tuning range of the RF frequencies in this study is so narrow that the influence of interference effects resulting from reflections off the walls of the metallic chamber can be neglected. The power of the RF radiation is adjusted by calibrated attenuators after the higher-harmonic generation. To detect RF transitions, we apply a state-selective field-ionization pulse \cite{peperPrecisionMeasurementIonization2019} and determine the ratio $\mathcal{Q}$ of ions detected in the time-of-flight-window corresponding to the selective ionization of 35$\,^2$D$_j$ over the total ion yield. The resulting spectra of the transition from $\Psi_e$ to the \molket{35}{2}{j'}{F'} asymptotes are shown in \autoref{fig:D_potentialandspectra}\,b), where panel III (IV) depicts the transfer to the \molket{35}{2}{j'}{1} (\molket{35}{2}{j'}{2}) asymptote. The spectra exhibit strong resonances at the atomic thresholds with a broad shoulder on the high-frequency side. The \molket{35}{2}{5/2}{2} asymptote forms an exception where no population transfer is observed. This can be understood in the two-state picture of Eq.~\eqref{eq:HFS:twostatemodel}. In $\Psi_e$, the $F=2$ character is correlated with $j=1/2$ so that the electric-dipole selection rule $\Delta j = 0, \pm 1$ prevents transfer to the $j'=5/2$ target state. Additional resonances in the gray shaded spectral regions (compare also \autoref{fig:D_potentialandspectra}\,a)) are attributed to bound-bound transitions.

\emph{Role of $\Pi$ states} We attribute the sharp resonances at threshold to photodissociation into the respective $\Pi$ continua. Because the electronic wavefunctions of $|m_l|=1$ states have a nodal plane containing the internuclear axis, the scattering interaction in these states tends to zero in first order. Driving a $\Sigma$ to $\Pi$ transition in a LRM thus selectively switches off the scattering interaction by promoting the Rydberg electron to a spatial orbit with near-zero overlap with the ground-state atom. At threshold, free pairs of atoms in the $\Pi$ continua have almost no kinetic energy (\autoref{fig:D_potentialandspectra}\,a)) and thus long de Broglie wavelengths which result in large Franck-Condon overlaps with the initial state $\Psi_e$. In contrast, the rates for photodissociation of $\Psi_e$ into the continuum of $\Sigma$ states do not exhibit sharp features at threshold and are strongly suppressed by the bound-continuum Franck-Condon factor. The photodissociation thus occurs almost exclusively into the continuum of $\Pi$ states, resulting in non-interacting photofragments.

The sequence of photoassociation of atoms in hyperfine state $F=1$ into an entangled molecular state and consecutive photodissociation at the threshold of an asymptote with the ground-state atom in hyperfine state $F'=2$ constitutes a remote quasi-spin flip of the ground-state atom, mediated and heralded through a Rydberg excitation. By additionally harvesting the strong, long-range interaction between Rydberg atoms, this approach could create long-range multiparticle entanglement of ground-state atoms \cite{durThreeQubitsCan2000,niederprumRydbergGroundState2016}. A reference measurement with molecules photoassociated in a state with pure triplet character (see Appendix~\ref{app:tripletdiss}) confirms that singlet-triple mixing is a prerequisite for the observation of a spin flip.

\section{Simulation of dissociation spectra}

We simulate the RF spectrum by calculating transition dipole moments, averaged over all accessible electronic states, and assuming an incoherent, irreversible population transfer due to fast dephasing of the continuum wavepackets. We calculate transition dipole moments between electronic states by evaluating the expectation value of the dipole operators $\mu_0$ and $\mu_\pm$ for parallel ($\Delta m_l=\Delta \Omega = 0$) and perpendicular ($\Delta m_l=\Delta \Omega = \pm 1$) transitions in the molecular frame, where angular matrix elements are evaluated analytically~\cite{varshalovichMATRIXELEMENTSIRREDUCIBLE1988} and radial matrix elements are calculated using Numerov's algorithm and model potentials~\cite{deiglmayrLongrangeInteractionsRydberg2016}. The spin-orbit interaction of the Rydberg electron mixes $m_l=0$ and $\abs{m_l}=1$. Thus all $\Sigma$ states correlated to asymptotes with $l>0$ also have contributions from $\abs{m_l}=1$. This contribution in principle allows for one-photon transitions from the states labeled $\Sigma$ here to $\Delta$ ($\abs{m_l}=2$) states, which have only recently been considered in the theoretical modeling of LRMs \cite{giannakeasDressedIonpairStates2020}.

The vibrational Franck-Condon factors are calculated through the modified Milne phase-amplitude method with continuum wavefunctions energy-normalized with respect to the corresponding molecular dissociation asymptote and uniform scattering phases\cite{sidkyPhaseamplitudeMethodCalculating1999a, peper2020heteronuclear}. The total transition rates are calculated in the molecular frame and neglect rotational H\"onl-London factors because the rotational states are experimentally not resolved. The resulting spectra are convoluted with the transform-limited Lorentzian line profile of the RF pulse. We account for the observed saturation of the RF transfer by applying an empirical saturation function of the form $A\left[1-\mathrm{exp}\left({-\gamma(\nu)/I_{\rm sat}}\right)\right]$ to the calculated spectrum $\gamma(\nu)$ and extract the two parameters $A$ and $I_{\rm sat}$ in a global least-squares analysis.
The result, depicted in \autoref{fig:D_potentialandspectra}\,b), reproduces the experimental features attributed to photodissociation transitions almost quantitatively, and features assigned to bound-bound transitions  qualitatively. We note that the modeling of bound-bound spectra depends strongly on the details of the computational method~\cite{peper2020heteronuclear} and we thus focus on the photodissociation resonances.

\begin{figure}[tb]
	\centering
	\includegraphics[width=\linewidth]{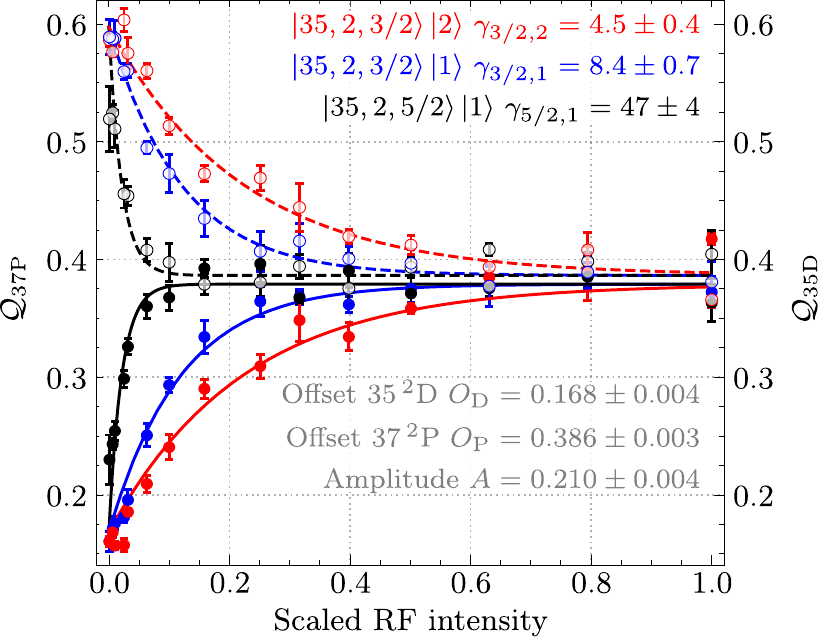}
	\caption{\label{fig:exp_dissrates} Fraction of ions $\mathcal{Q}$ detected in the time-of-flight window set for 37$\,^2$P$_j$ (open symbols, left axis) and 35$\,^2$D$_j$ (filled symbols, right axis) as a function of the applied RF power at the frequencies 86.6941 GHz (black), 86.6669 GHz (blue), and 86.2052 GHz (red), corresponding to the transitions from the molecular state to the respective dissociation thresholds specified in the inset. RF powers are given as a fraction $I_\mathrm{rel}$ of the maximal applied power. Also shown are the global fit (solid and dashed lines for 37$\,^2$P$_j$ and 35$\,^2$D$_j$, respectively), fit parameters and statistical uncertainties of the theoretical model to the combined experimental data (see text for details).}
\end{figure}

\section{State tomography}
A closer inspection of the calculated photodissociation rates reveals for some transitions a strong dependence on $\Omega$ of the initial state $\Psi_e$ (\autoref{tab:HFS:TheoRates}). Because of negligible spin-orbit coupling in the electron-K scattering, molecular states of different $\Omega$ are degenerate~\cite{deissObservationSpinorbitdependentElectron2020,peper2020heteronuclear} and not resolved in photoassociation. We employ photodissociation to determine the distribution of $\Omega$ in the photoassociated molecular sample by comparing measured and calculated photodissociation rates. The rates are measured at the respective dissociation thresholds by varying the RF power $I_\mathrm{rel}$ using a calibrated RF attenuator and recording the fraction $\mathcal{Q}$ of ions detected in the time-of-flight windows set for $37\,^2$P and $35\,^2$D atoms. The results obtained for different attenuations of the RF power are depicted in \autoref{fig:exp_dissrates}. We extract dissociation rates by a global fit of the kinetic model

\begin{align*}\label{eq:HFS:FitModelRates}
\mathcal{Q}_{{\rm 37P},\kappa} & = A \exp\left[-\gamma_\kappa  I_\mathrm{rel}\right]  + O_\mathrm{P} \\
\mathcal{Q}_{{\rm 35D},\kappa} & = A \left(1-\exp\left[-\gamma_\kappa  I_\mathrm{rel}\right]\right)  + O_\mathrm{D}
\end{align*}
of an irreversible reaction of first order to the data (see~\autoref{fig:exp_dissrates}), where $\gamma_\kappa$ ($\kappa=\{j,F\}$) are the asymptote-dependent dissociation rates, $A$ represents the fraction of the $37\,^2$P signal originating from LRMs in state $\Psi_e$, and $O_\mathrm{P}$ and $O_\mathrm{D}$ are experimental offsets resulting from from off-resonant excitation of isolated Rydberg atoms or Rydberg-atom pairs~\cite{deiglmayrLongrangeInteractionsRydberg2016}.

\begin{table}[t!]
	\caption{\label{tab:HFS:TheoRates}Theoretical and experimental photodissociation rates relative to the rate for the transition to the \molket{35}{2}{3/2}{1} asymptote. The theoretical ratios of rates are $\Omega$ dependent, from which $\Omega$-averaged ratios are determined by a fit to the experimental data.}
	\begin{center}
		\begin{tabular}{c S[table-format=4.4] S[table-format=4.4] S[table-format=4.4] c}
			\hline\hline
				& \multicolumn{4}{c}{Ratio over \molket{35}{2}{3/2}{1}}\\
			\cmidrule{2-5}
			 		&  \multicolumn{2}{c}{Theo. ($\Omega$)} & \multicolumn{1}{c}{Theo.} &\multirow{2}{*}{Exp.} \\
			 		  \multicolumn{1}{c}{}    & \multicolumn{1}{c}{$1/2$}   &  \multicolumn{1}{c}{$3/2$} & \multicolumn{1}{c}{($\Omega$-aver.)} &         \\ \hline
			\molket{35}{2}{5/2}{1}  &     6.06    &  5.94 & 5.97 & $5.6\pm0.7$\\
			\molket{35}{2}{3/2}{2} &      1.04     & 0.38 & 0.54 & $0.54\pm0.06$\\
			\molket{35}{2}{5/2}{2} &  0.003     & 0.002 & 0.002    & $0.0\pm0.1$\\
			\hline\hline
		\end{tabular}
	\end{center}
\end{table}

For comparison with theory, ratios of photodissociation rates are given in \autoref{tab:HFS:TheoRates}, canceling the unknown absolute intensity of the RF radiation. Numerical simulations show that for comparable initial populations and transfer rates of the $\Omega$ states, the experimentally extracted rate is close to the weighted mean of the transfer rates. We thus determine the relative initial populations in $\Omega=1/2$ and $3/2$ by minimizing the deviation between the experimental ratios and a weighted mean of the theoretical ratios from \autoref{tab:HFS:TheoRates}, yielding a fraction of \SI{68 \pm 6}{\percent} of initial molecules in the $\Omega=3/2$ state.

\begin{figure}[tb]
	\centering
    \includegraphics[width=\linewidth]{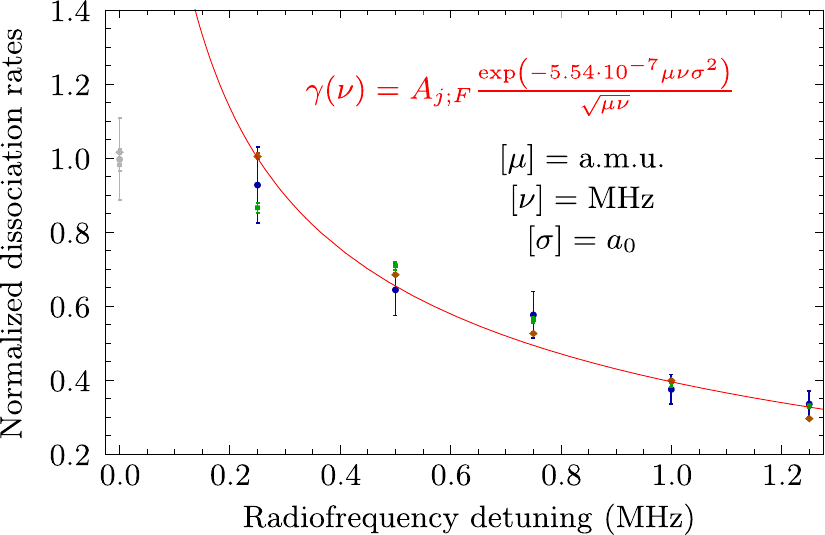}
	\caption{\label{fig:energydependencedissrrates}Energy-dependence of photodissociation rates from the $\Psi_e$ state to the \molket{35}{2}{3/2}{1} (green), \molket{35}{2}{5/2}{1} (blue), and \molket{35}{2}{3/2}{2} (brown) asymptotes obtained from fits to power-dependent measurements at different RF detunings above the dissociation threshold. The error bars indicate the standard deviation of the fitted rates. The simple model explained in the text predicts the energy dependence $\gamma(E_k)$ given on the figure. Here, $A_{j;F}$ is an asymptote-dependent amplitude, $\mu$ the reduced mass of the molecule in atomic mass units, $\nu=E_k/h$ the detuning from threshold in megahertz, and $\sigma$ the Gaussian width of the initial molecular state in Bohr radii. The red curve is a global fit of this model to the data points, weighted by their standard deviations, with free parameters $A_{3/2;1}$, $A_{3/2;2}$, $A_{5/2;1}$, and $\sigma$. The shown photodissociation rates are normalized by the fitted amplitudes $A_{j;F}$. The rates at \SI{0}{MHz} RF detuning (gray data points) were not included into the fit because of the divergence of the theoretical model at this point.}
\end{figure}

The dependence of the photodissociation rate on the excess kinetic energy $E_k$, controlled by the RF detuning $\nu$ from threshold, yields additional information on the vibrational wavefunction of the initial state $\Psi_e$: as the de Broglie wavelength of the continuum wavefunction becomes shorter than the extent of the initial near-Gaussian wavefunction (\autoref{fig:P_potentialandspectra}), their mutual overlap  averages to zero. To study the energy-dependence of the photodissociation rates above threshold in more detail, we repeat the measurement and analysis of photodissociation rates described above at several values of the RF detuning. The results are shown in \autoref{fig:energydependencedissrrates}, normalized to the global fit of the model function shown on the figure. The model function is obtained by analytically solving the overlap integral of a Gaussian wave function with width $\sigma$ and a $s$-wave continuum wavefunction of the form $1/\sqrt{k}\sin(kR)$, where $k$ is the magnitude of the continuum wavevector and $R$ the interatomic separation, and extracting the envelope of the resulting oscillating function~\cite{chinRadiofrequencyTransitionsWeakly2005}. This yields an energy dependence of the dissocation rate $\gamma$ of the form $\gamma(E_k)=\exp\left(- E_k \sigma^2\right)/\sqrt{E_k}$~\cite{chinRadiofrequencyTransitionsWeakly2005}. A fit of this model function to the measured rates reproduces the experimental observations and yields a spatial width of the vibrational wavefunction of $\Psi_e$ $\sigma=\SI{170\pm80}{a_0}$ where the reported uncertainty is the statistical uncertainty of the nonlinear regression. This result agrees with the value of \SI{89}{a_0} extracted from a fit of a Gaussian to the calculated vibrational wavefunction.

\section{Conclusion and outlook}
We demonstrated the use of photodissociation spectroscopy of LRMs for the characterization of their properties by projection onto an unperturbed atomic basis and comparison to theoretical calculations. Further, we presented a scheme for a complete and remote flip of the hyperfine state of free ground-state atoms through the interaction with a Rydberg atom, where we have exploited the near-degeneracy of fine- and hyperfine structure in \potassium{} $37\,^2{\rm P}_j$ states. Similar degeneracies exist in other systems, \textit{e.g.} in $^{87}$Rb ($26\,^2{\rm P}_j$) and heteronuclear LRMs ($^{39}$K$^{133}$Cs, Cs(53\,$^2{\rm D}_j$))~\cite{eilesFormationLongrangeRydberg2018,peper2020heteronuclear}. In the future, the vibrational wavefunction of a LRM might be characterized more precisely by photodissociation to a repulsive PEC \cite{shapiroPhotofragmentationMappingNuclear1981,scotthopkinsCommunicationImagingWavefunctions2011,schmidtSpatialImagingVibrational2012}. Photodissociation of oriented LRMs \cite{Krupp2014,engelPrecisionSpectroscopyNegativeIon2019} might be used to create counter-propagating beams of atoms with well-controlled kinetic energy and entangled internal quantum states.\\

\begin{acknowledgments}
We thank Fr\'ed\'eric Merkt for continuous support and invaluable discussions. This work was supported by the ETH Research Grant ETH-22 15-1 and the NCCR QSIT of the Swiss National Science Foundation.
\end{acknowledgments}

\appendix

\section{Calculation of photoassociation intensities and mixing coefficients}
\label{app:calcPA}

To calculate the relative strength of the photoassociation resonances (represented by the height of the bars in Fig.~2\,b)) we assume that there is no fixed phase relation of the colliding ground-state atoms over the range of the vibrational wavefunction of the photoassociated LRM. This assumption is reasonable, because the thermal de~Broglie wavelength of \potassium{} atoms at \SI{20}{\micro\kelvin} is about $1000\,a_0$, much shorter than the internuclear separations considered here, and several partial waves contribute to the initial scattering wavefunction at the equilibrium distance of the photoassociated molecules $R_e \sim 2350\,a_0$. For a homogeneous gas of atoms initially prepared in the hyperfine state $F$, the relative photoassociation rates were consequently estimated through the expression
\begin{equation}
\Gamma_{\kappa,\nu}^F = \gamma^{\rm PA}_{\kappa} \sum_j f_{j} \int R^2 \left|\Psi_{\kappa,\nu}(R)\right|^2 \left|\alpha_{j;F}^\kappa(R) \right|^2\mathrm{d}R,
\end{equation}
where $\kappa$ labels the electronic state addressed in photoassociation and $\Psi_{\kappa,\nu}$ is the vibrational wavefunction of level $\nu$ (see Supplementary Material of Ref.~\cite{peper2020heteronuclear}). The global coefficients $\gamma^{\rm PA}_{\kappa}$ contain \textit{e.g.} electronic transition dipole moments and photoassociation intensities and were determined by comparison with the experimental spectra, while the ratio of the atomic $37\,^2\mathrm{P}_{j}\leftarrow 4\,^2\mathrm{S}_{1/2}$ transition dipole moments was determined experimentally to be $\mathcal{F}=f_{j=3/2}/f_{j=1/2}=\num{3.3(5)}$. The deviation from the statistically expected ratio of 2:1 results from a Cooper minimum in the photoionization cross section \cite{huangOscillatorStrengthPrincipal1981}. The absolute squares of the mixing coefficients $\alpha_{j;F}^\kappa$ are obtained by the expression
\begin{equation}
   \left|\alpha_{j;F}^\kappa(R)\right|^2 = \sum_{m_j,m_F} \left|\braket{37,1,j,m_j,F,m_F|\phi_{\kappa}(R)}\right|^2,
\end{equation}
where $\phi_{\kappa}(R)$ is the electronic wavefunction of the photoassociated state.

\section{Characterization of a state with pure $F$ character}
\label{app:tripletdiss}
\begin{figure}[t!]
	\centering
    \includegraphics[width=\linewidth]{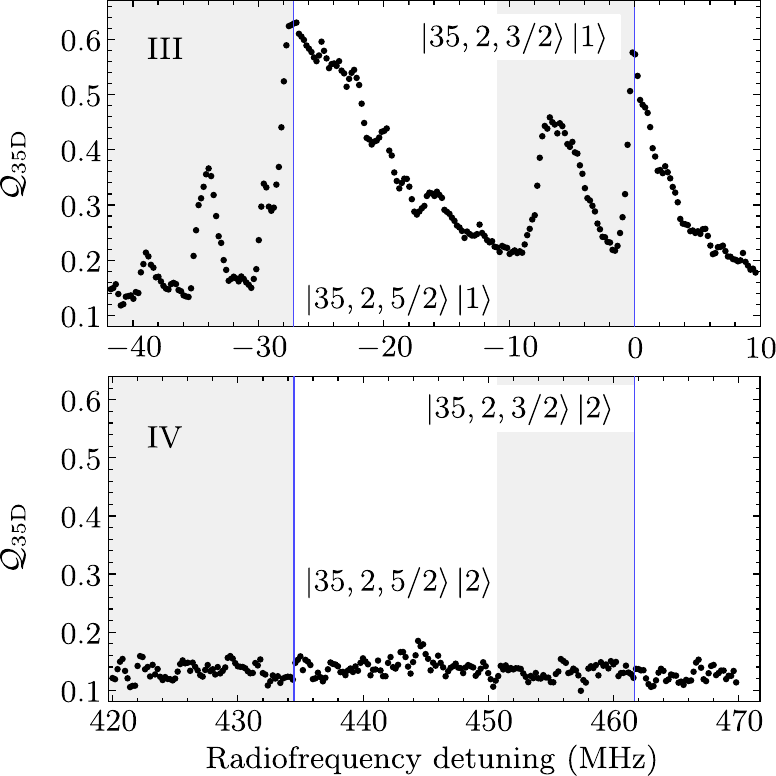}
	\caption{\label{fig:D_tripletspectra} RF spectra of the $^{3}\Sigma^+$ ($v=0$) K$_2$ molecules (dark green bar in the upper panel of Fig.~2\,b)) in the vicinity of the transition to the spin-orbit-split  (III) \molket{35}{2}{j'}{1} and (IV) \molket{35}{2}{j'}{2} asymptotes after a \SI{30}{\micro\second} photoassociation pulse followed by a \SI{10}{\micro\second} RF pulse. Zero RF detuning is set to \SI{86.662}{GHz}, \SI{16.31}{MHz} below the atomic transition $35\,^2\mathrm{D}_{3/2} \leftarrow 37\,^2\mathrm{P}_{3/2}$, corresponding to the transition from the bound molecular state to the dissociation continuum associated with the \molket{35}{2}{3/2}{1} asymptote. Gray-shaded areas mark regions where transitions to bound states are possible. The vertical thin blue lines mark the transition frequencies from the initial molecular state to the indicated dissociation asymptotes.}
\end{figure}

For completeness, we also present a photodissociation spectrum of molecules photoassociated in the $v=0$ level of the $^3\Sigma^+$ state below the \molket{37}{1}{3/2}{1} asymptote. The resulting RF spectra close to the \molket{35}{2}{j'}{1} (III) and \molket{35}{2}{j'}{2} (IV) asymptotes are depicted in \autoref{fig:D_tripletspectra}. The RF spectrum in the vicinity of the \molket{35}{2}{j'}{1} asymptotes is qualitatively very similar to the RF spectrum obtained for the entangled $\Psi_e$ state, containing features assigned to bound-bound and bound-continuum transitions. The RF spectrum in the vicinity of the \molket{35}{2}{j'}{2} asymptotes, however, does not show any population transfer, confirming the pure \molket{37}{1}{3/2}{1} character of the unentangled molecular $^3\Sigma^+$ state.

\begin{table}[t!]
	\caption{\label{tab:HFS:TheoRatesApp} Calculated photodissociation rates relative to the rate for the transition to the \molket{35}{2}{3/2}{1} asymptote. Note the difference with respect to Table~I, where the ratios are normalized separately for each value of $\Omega$ for better comparison. The different approximations are described in the text.}	
	\begin{center}
		\begin{tabular}{cccccc}
			\hline\hline
				& \multicolumn{5}{c}{Ratio over \molket{35}{2}{3/2}{1}}($\Omega=1/2$)\hfill\\
			\cmidrule{2-6}
			 		& \multirow{2}{*}{atomic}  & \multicolumn{2}{c}{electronic}& \multicolumn{2}{c}{vibronic} \\
			 		&     &    $\Omega=1/2$     &       $\Omega=3/2$         & $\Omega=1/2$   &  $\Omega=3/2$      \\ \hline
			\molket{35}{2}{3/2}{1} &  1    & 1     &  1.5  & 1    & 1.53  \\
			\molket{35}{2}{5/2}{1} &  9    & 7.68  & 11.5  & 6.06 & 9.06  \\
			\molket{35}{2}{3/2}{2} &  1.0  & 1.20  &  0.601  & 1.05 & 0.587  \\
			\molket{35}{2}{5/2}{2} &  0    & 0.003 &  0.002 & 0.0035 & 0.0027 \\
			\hline\hline
		\end{tabular}
	\end{center}
\end{table}

\section{Simpler models for the radiofrequency photodissociation spectra}
\label{app:simPD}

In \autoref{tab:HFS:TheoRatesApp}, we compare the results of the full simulation of threshold photodissociation rates, described in the article and reproduced in the column ``vibronic'', to two simpler models. In the ``atomic'' approximation, only the relative transition strengths of the basis states in the two-state approximation (\autoref{eq:HFS:twostatemodel}) are evaluated via the Wigner-Eckart theorem \cite{varshalovichMATRIXELEMENTSIRREDUCIBLE1988} and the mean value of the mixing coefficient $|\alpha_{1/2;2}^e|^2=0.168$, which does not depend on $\Omega$. To refine this estimate, we calculate the expectation values of the dipole moments for transitions to $\Pi$ states with respect to the initial vibrational level $\Psi_e$. The resulting rates capture the relative strengths of the observed transitions, but deviate still significantly from the experimental values. The ``vibronic'' calculation predicts  overall smaller ratios, which we attribute to the averaging of the diverging photodissociation rates at threshold and, to a lesser extent, to the inclusion of transitions into the continuum of $\Sigma$ states.

\end{document}